\documentclass[%
aip,
amsmath,amssymb,
reprint,%
]{revtex4-1}

\usepackage{graphicx}
\usepackage{dcolumn}
\usepackage{bm}

\usepackage[utf8]{inputenc}
\usepackage[T1]{fontenc}
\usepackage{mathptmx}
\usepackage{etoolbox}

\makeatletter
\def\@email#1#2{%
	\endgroup
	\patchcmd{\titleblock@produce}
	{\frontmatter@RRAPformat}
	{\frontmatter@RRAPformat{\produce@RRAP{*#1\href{mailto:#2}{#2}}}\frontmatter@RRAPformat}
	{}{}
}%
\makeatother
\begin{document}
	
	\preprint{AIP/123-QED}
	
	\title[]{Fast reversal of N\'{e}el vectors in antiferromagnets via domain-wall motion driven by vertically injected spin currents}
	\author{Jiaxin Du}
	\affiliation{College of Physics and Hebei Advanced Thin Films Laboratory, Hebei Normal University, Shijiazhuang 050024, People's Republic of China}
	\author{Mei Li}
	\affiliation{College of Physics Science and Technology, Yangzhou University, Yangzhou 225002, People's Republic of China}
	\author{Bin Xi}
	\affiliation{College of Physics Science and Technology, Yangzhou University, Yangzhou 225002, People's Republic of China}
	\author{Xue Zhang}
	\affiliation{College of Physics and Hebei Advanced Thin Films Laboratory, Hebei Normal University, Shijiazhuang 050024, People's Republic of China}
	\author{Yi Sun}
	\affiliation{School of Electronic Engineering, Jiangsu Vocational College of Electronics and Information, Huaian 223003, People's Republic of China}
	\author{Chun-Gui Duan}
	\affiliation{College of Physics and Hebei Advanced Thin Films Laboratory, Hebei Normal University, Shijiazhuang 050024, People's Republic of China}
	\affiliation{Authors to whom any correspondence should be addressed.}
	\author{Jie Lu}
	\affiliation{College of Physics Science and Technology, Yangzhou University, Yangzhou 225002, People's Republic of China}
	\affiliation{Authors to whom any correspondence should be addressed.}
	\email{duancg@hebtu.edu.cn, lujie@yzu.edu.cn}

	\date{\today}
	
	\begin{abstract}
		In this work, we investigate the steady propagation of 180-degree domain walls (180DWs) of N\'{e}el vectors in
		thin antiferromagnetic strips under perpendicularly injected spin currents with various polarization orientations.
		Our results show that only spin currents polarized normally to the strip plane 
		guarantee a quick and steady rigid flow of 180DWs, 
		thus realize a fast reversal of N\'{e}el vectors in antiferromagnets (AFMs).		
		Different from the common ``current-in-plane'' geometry which is feasible only for metallic AFMs, 
		our ``current-out-of-plane'' layout under investigation can further apply to insulating AFMs (which are more common 
		in real applications) via quantum tunneling effect.
		Results from this work paves the way for fine control of  N\'{e}el vectors in (both metallic and insulating) AFM strips
		and further development of novel magnetic nanodevices based on them. 
	\end{abstract}
	
	\maketitle

	Fast reversal of N\'{e}el vector in antiferromagnets (AFMs) is crucial for the realization of various advanced
	magnetic nanodevices\cite{Tserkovnyak_RMP_2018,XRWang_EPL_2020,HaimingYu_PhysRep_2021,LiuZhiqi_AM_2024}, such as Terahertz nanooscillators\cite{RCheng_PRL_2016,Chirac_PRB_2020}, 
	high-performance magnetic random access memories\cite{Shadman_APL_2019}, 
	and high-sensitivity Hall devices\cite{Fischer_PRApp_2020}, etc. 
	In most existing treatments\cite{RCheng_PRL_2016,Chirac_PRB_2020,Shadman_APL_2019,Fischer_PRApp_2020,HaoWu_NC_2022,Weissenhofer_PRB_2023,SongCheng_SciAdv_2024}, AFM films or strips are considered to be single-domain 
	in N\'{e}el vectors	due to the much larger exchange coupling in AFMs compared with that in ferromagnets (FMs).
	However, for AFMs with large-enough film or long-strip geometries, the flipping of N\'{e}el vectors is generally 
	not realized by the coherent reversal but by the motion of antiferromagnetic 180-degree domain walls (AFM-180DWs), 
	which can be driven by several stimuli, such as staggered fields\cite{XRWang_PRB_2018}, 
	spin waves (magnons)\cite{Tchernyshyov_PRB_2014,Brataas_PRB_2018}, and in-plane currents\cite{NiuQian_PRB_2014,Sinova_PRB_2016,Rodrigues_PRB_2017,Mondal_PRB_2018,Fujimoto_PRB_2021,XRWang_PRB_2024}, etc.
    Among them, in-plane currents are of the most interest due to their
    simplicity in device-structure design, independence of injecting electrons' spin polarity, low total current intensity, and so on.
    However, all these advantages rely on the metallicity of AFM materials. 
    In reality most AFMs with excellent properties are insulators or semiconductors.
    Therefore the aforementioned ``current-in-plane'' geometry may not be the best choice.
    
	Inspired by the early works in FM spin-valves\cite{Rebei_Mryasov_PRB_2006,Khvalkovskiy_PRL_2009,Boone_PRL_2010,PBHe_EPJB_2013,jlu_PRB_2019,jlu_PRB_2021,jlu_NJP_2023}, AFM-180DWs can also be driven by spin currents coming from a perpendicular electron flow 
	that polarized by an adjacent FM layer (the ``polarizer'').
	Without loss of generality, we only consider the effects of transmitted (through the polarizer) electrons 
	hence the spin current propagates from FM to 
	AFM layer meantime has the polarization direction parallel to the magnetization orientation of the polarizer\cite{Ralph_JMMM_2008,Gomonay_LowTempPhys_2014,KWKim_IEEENano_2020}.
	On the other hand, the AFM strip can be either metallic or insulting in which the electron current achieves 
	stable existence through direct conduction or quantum tunneling.
	In fact, this geometry (so-called ``discrete FM/AFM heterostructure'') has already been proposed for decades\cite{MacDonald_PRL_2008,ZhenWei_JAP_2009,Haney_PRB_2014,Gomonay_LowTempPhys_2008,Gomonay_PRB_2010,Gomonay_PRB_2012,KWKim_APL_2017,KWKim_PRB_2021}, 
	but the focus has been on the coherent reversal of N\'{e}el vectors in it, 
	rather than the reversal through the motion of AFM-180DWs.
	
	In this Letter, by transforming the discrete AFM Heisenberg model to a more convenient continuum model
	in nanomagnetism, as well as introducing adequately Gilbert damping and Slonczewski-torque-induced energy gain,
	we systematically investigate the dynamics of AFM-180DWs driven by vertically-injected spin currents in thin AFM strips.
	We focus on the rigid-flow mode of walls (especially its applicable range and stability) under three typical choices 
	of polarizer orientation.
	It turns out that only perpendicular polarizers induce steady rigid flows of AFM-180DWs with high wall velocity,
	thus realize ultra-fast reversal of N\'{e}el vectors.

\begin{figure}
	\includegraphics[width=0.35\textwidth]{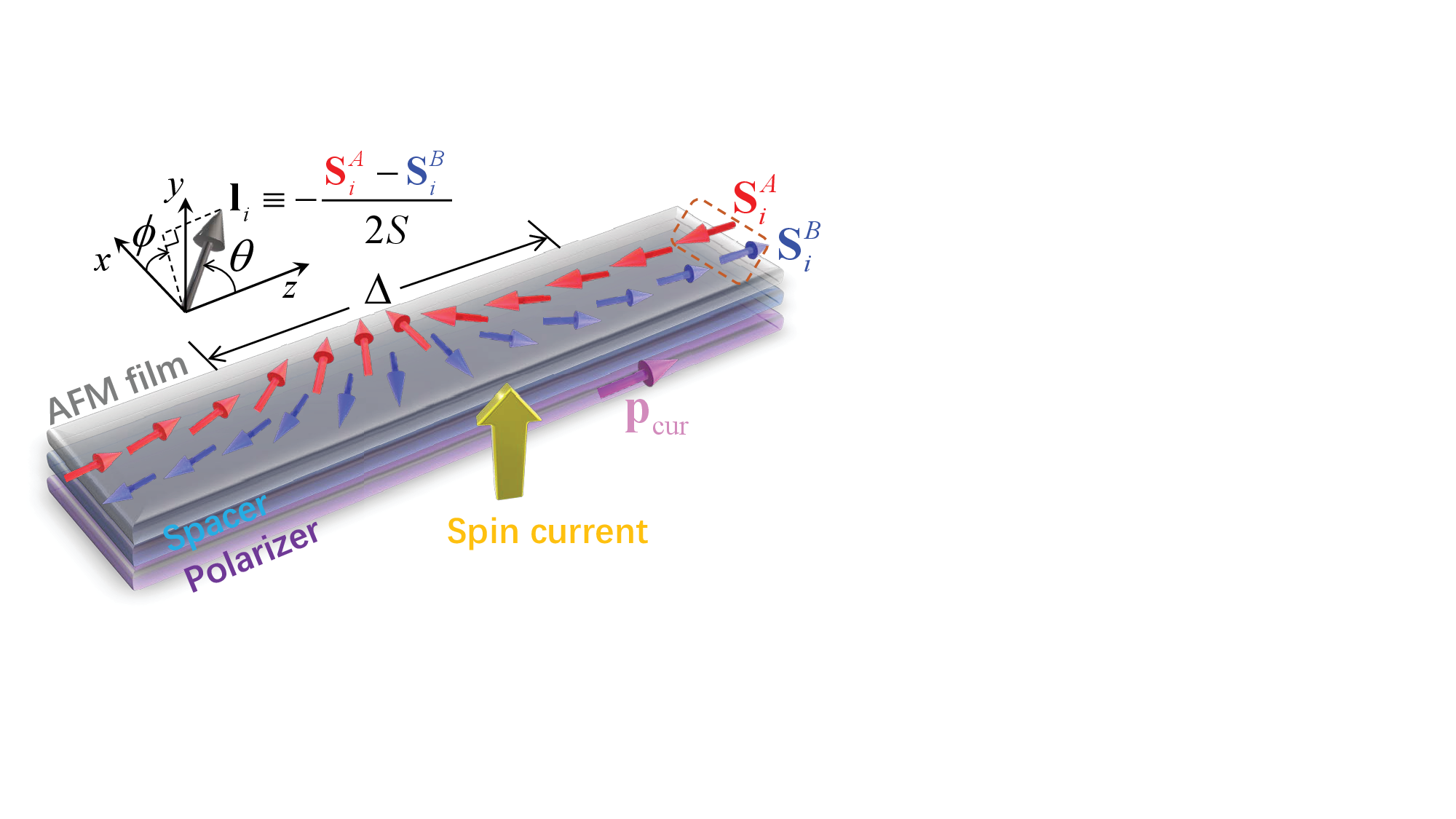}
	\caption{\label{fig1} Schematic illustration of an AFM strip and a FM polarizer, separated by a metallic spacer.
		The AFM strip has a length $L_z$, width $w$ and thickness $d$ ($L_z\gg w\gg d$).
		It is modeled by a 1D AFM Heisenberg spin chain, with $\mathbf{S}_i^A$ and $\mathbf{S}_i^B$ 
		(both of magnitude $S$) constituting the unit cell. The corresponding N\'{e}el vector is $\mathbf{l}_i$.
		An electron flow first passes through the polarizer (with unit magnetization $\mathbf{p}_{\mathrm{cur}}$)
		then vertically injects onto the AFM strip, carrying a spin current with polarization along
		$\mathbf{p}_{\mathrm{cur}}$. Driven by the resulting Slonczewski spin-transfer torque, an AFM-180DW can
		propagate in $z-$direction.}
    \end{figure}

	The system setup is illustrated in Fig. \ref{fig1}, in which an AFM strip extends in $z$ axis with length $L_z$,
	meantime keeps a finite width $w$ in $x$ direction and a thin thickness $d$ along $y$ axis. 
	This AFM material can be either metallic or insulating, as long as $d$ is small enough (a few or tens of nanometers)
	so that the itinerant electrons have a considerable probability of passing through it. 
	A single-domain metallic FM layer with unit magnetization $\mathbf{p}_{\mathrm{cur}}$ 
	acts as a polarizer that turns the spins of itinerant electrons to follow the orientation of its magnetization.
	A charge current with density $j_{\mathrm{cur}}$ runs from the AFM to FM layers, leading to a reversed 
	electron flow with the same strength.
	It then carries a spin current perpendicularly injecting onto the AFM layer with polarization parallel to $\mathbf{p}_{\mathrm{cur}}$.
	Between the AFM strip and polarizer, a normal metal (spacer) with adequate thickness is sandwiched to avoid 
	the magnetic proximity effect meantime prevent spin dephasing.

	We start from the classical Heisenberg Hamiltonian including the exchange coupling between classical spin vectors 
	on a one dimensional (1D) lattice with $2N$ sites. Together with the biaxial anisotropy, the Hamiltonian reads
	\begin{equation}\label{HeisenbergHamiltonianDiscrete}
		H=J\sum^{2N-1}_{\alpha=1}\mathbf{S}_{\alpha}\cdot\mathbf{S}_{\alpha+1}
		                        +\sum^{2N}_{\alpha=1}\left[K_y\left(\mathbf{S}_{\alpha}\cdot \mathbf{e}_y\right)^2-K_z\left(\mathbf{S}_{\alpha}\cdot \mathbf{e}_z\right)^2\right]
	\end{equation}
	where $\mathbf{S}_{\alpha}$ is the spin vector on the $\alpha-$th site ($|\mathbf{S}_{\alpha}|\equiv S$), $J>0$ is the positive exchange 
	energy resulting in an AFM ground state, and $K_{z(y)}$ denotes the easy (hard) anisotropy in $z$ ($y$) axis.
	
	For a 1D AFM chain with $2N$ sites and bond length $c$, it is bipartite 
	thus the length of unit cell (labeled by $i=1,\cdots, N$) becomes $a_0=2c$ 
	and the sublattice indices are denoted as $A$ and $B$. 
	Correspondingly, $L_z=Na_0$.
	Then we define the discrete magnetization and staggered N\'{e}el vectors as
	\begin{equation}\label{m_l_definitions}
		\mathbf{m}_{i}\equiv -\frac{\mathbf{S}_{i}^{A}+\mathbf{S}_{i}^{B}}{2 S},\qquad \mathbf{l}_{i}\equiv -\frac{\mathbf{S}_{i}^{A}-\mathbf{S}_{i}^{B}}{2 S},
	\end{equation}
	so that $\mathbf{m}_{i}^2+\mathbf{l}_{i}^2\equiv 1$ and $\mathbf{m}_{i}\cdot\mathbf{l}_{i}\equiv 0$.
	The minus signs come from the fact that magnetic moments and spins of electrons are opposite.
	By taking: $a_0\sum_{i=1}^{N}\rightarrow \int_{0}^{L_z}dz$,
	$\mathbf{m}_i\rightarrow\mathbf{m}(z,t)$, $\mathbf{l}_i\rightarrow\mathbf{l}(z,t)$,
	$\mathbf{m}_{i+1}-\mathbf{m}_{i}\rightarrow a_0\partial\mathbf{m}/\partial z$,
	$\mathbf{l}_{i+1}-\mathbf{l}_{i}\rightarrow a_0\partial\mathbf{l}/\partial z$,
	and  introducing the normalized N\'{e}el vector field $\mathbf{n}(z,t)\equiv\mathbf{l}(z,t)/|\mathbf{l}(z,t)|$,
	Eq. (\ref{HeisenbergHamiltonianDiscrete}) is transformed to its 
	continuum counterpart via the ``Hamiltonian approach''\cite{Brataas_PRB_2016,Stamps_PRB_2021}
	as $E_{\mathrm{AFM}}=\int_{0}^{L_z}\frac{dz}{a_0}\mathcal{E}$ with
	\begin{equation}\label{HeisenbergHamiltonianContinuum}
		\mathcal{E}\equiv  \frac{a}{2}|\mathbf{m}|^2+\frac{A}{2}\left|\frac{\partial\mathbf{n}}{\partial z}\right|^2+ L\left(\mathbf{m}\cdot\frac{\partial\mathbf{n}}{\partial z}\right)+\frac{k_y}{2}l_y^2-\frac{k_z}{2}l_z^2,
	\end{equation}
	where $a=8JS^2$, $A=a_0^2 J S^2$, $L=2a_0 JS^2$, and $k_{y(z)}=4K_{y(z)}S^2$.
	In the deduction of Eq. (\ref{HeisenbergHamiltonianContinuum}), the following approximations
	have been made:
	(i) The marginal term, $-\frac{1}{2}JS^2[(\mathbf{m}_1^2-\mathbf{m}_N^2)-(\mathbf{l}_1^2-\mathbf{l}_N^2)]$,
	has been dropped off, 
	(ii) The typical length scales that $\mathbf{m}$ and $\mathbf{n}$ vary are much longer than $a_0$,
	(iii) The existence of magnetic anisotropy prohibits the occurrence of spin flop under weak
	external stimuli, thus $|\mathbf{m}|\ll |\mathbf{n}|\equiv1$.
	
	Next, the kinetic energy of this bipartite AFM in the continuum approximation reads\cite{Brataas_PRB_2016}
	\begin{equation}\label{K_definition}
		K=\int_{0}^{L_z}\frac{dz}{a_0}\mathcal{K},\quad \mathcal{K}=\rho\mathbf{m}\cdot\left(\frac{\partial\mathbf{n}}{\partial t}\times\mathbf{n}\right),
	\end{equation}
	with $\rho\equiv 2 S\hbar$. It comes from the expanding of spin-pair
	Berry phase in small deviations from the antiparallel configuration ($\mathbf{m}\ll 1$),
	and discarding terms of the order $|\mathbf{m}|^4$ and higher. 
	The Lagrangian is then 
	constructed	as $L=\int_{0}^{L_z}\frac{dz}{a_0}\mathcal{L}$ with $\mathcal{L}\equiv \mathcal{K}-\mathcal{E}$.
	By minimizing it with respect to $\mathbf{m}$, we obtain
	\begin{equation}\label{m_slave}
		\mathbf{m}=\frac{\rho}{a}\left(\frac{\partial\mathbf{n}}{\partial t}\times\mathbf{n}\right)-\frac{L}{a}\left(\frac{\partial\mathbf{n}}{\partial z}\right),
	\end{equation}
	implying that $\mathbf{m}$ is indeed a slave variable of $\mathbf{n}$.
	
	Putting Eq. (\ref{m_slave}) back into Eqs. (\ref{HeisenbergHamiltonianContinuum}) and (\ref{K_definition}),
	the Lagrangian is fully described by the N\'{e}el vector and its temporal and spatial derivatives.
	Then we introduce the 1D unit cell with volume $c\times c\times 2c=a_0^3/4$, 
	and the effective saturation magnetization
	$M_s\equiv g_e S \mu_{\mathrm{B}}/(a_0^3/4)$ with $g_e\approx 2$ being the Land\'{e} $g$-factor of 
	electrons and $\mu_{\mathrm{B}}$ indicating the Bohr magneton. 
	By noting $L^2\equiv aA/2$, the Lagrangian density finally becomes
	\begin{eqnarray}\label{Lagrangian_Dimensionless}
		\frac{\mathcal{L}_{\mathrm{AFM}}\left[\mathbf{n},\partial_{Z}\mathbf{n},\partial_{\tau}\mathbf{n}\right]}{JS^2/(\mu_0 M_s^2 a_0^3)}&=&\left(\frac{\partial \mathbf{n}}{\partial\tau}\right)^2-2\left(\frac{\partial \mathbf{n}}{\partial\tau}\times\mathbf{n}\right)\cdot\frac{\partial \mathbf{n}}{\partial Z} \nonumber \\ 
		  & & \  -\left(\frac{\partial \mathbf{n}}{\partial Z}\right)^2-\kappa_y n_y^2+\kappa_z n_z^2,
	\end{eqnarray}
	in which $\mu_0$ is the vacuum permeability, $\kappa_{y(z)}=8K_{y(z)}/J$, 
	$Z=z/a_0$ and $\tau=t/t_0$ are the dimensionless spacial and temporal coordinates. 
	Here we have defined $t_0\equiv(\gamma_0 H_E)^{-1}$, where 
	$\gamma_0=\mu_0\gamma_e$ with $\gamma_e$ being the electron gyromagnetic ratio, 
	and	$H_E=JS/(\gamma_0\hbar)$ indicates the effective AFM exchange field.
	In addition, the Rayleigh dissipative functional including the Gilbert damping and 
	the Slonczewski spin-transfer term from the vertically injected spin current reads\cite{Gomonay_LowTempPhys_2014,KWKim_IEEENano_2020,Gomonay_PRB_2010,Gomonay_PRB_2012,KWKim_APL_2017,KWKim_PRB_2021}
	\begin{equation}\label{RayleighFunctional}
		\mathcal{R}_{\mathrm{AFM}}=\frac{\alpha_{\mathrm{G}}}{4\gamma_0 M_s t_0^2}\left(\frac{\partial\mathbf{n}}{\partial\tau}\right)^2 + \frac{\epsilon\hbar j_{\mathrm{cur}}}{4d e \mu_0 M_s^2 t_0}\frac{\partial\mathbf{n}}{\partial\tau}\cdot\left(\mathbf{n}\times\mathbf{p}_{\mathrm{cur}}\right),
	\end{equation}
	where $\alpha_{\mathrm{G}}$ is the damping coefficient, $\epsilon$ ($0<\epsilon<1$) is the polarization efficiency, and
	$e(>0)$ is the absolute value of electron charge.
	Now the magnetization dynamics can be well-described by the ``Euler-Lagrange-Rayleigh'' equation
	\begin{equation}\label{EulerLagrangeRayleighEquation}
		\partial_{\mu}\left(\frac{\partial\mathcal{L}_{\mathrm{AFM}}}{\partial(\partial_{\mu} \Lambda)}\right)-\frac{\partial\mathcal{L}_{\mathrm{AFM}}}{\partial \Lambda} +\frac{\partial\mathcal{R}_{\mathrm{AFM}}}{\partial(\partial_{\tau} \Lambda)}=0,
	\end{equation}
	in which $\Lambda$ denotes any generalized coordinate. 

	Next, $\mathbf{n}(Z,\tau)$ is fully described by its polar and azimuthal angles: $[\theta(Z,\tau),\phi(Z,\tau)]$.
	We then expand the configuration space of AFM-180DWs by the generalized Walker ansatz,
	\begin{equation}\label{WalerAnsatz}
		\ln\tan\frac{\vartheta(Z,\tau)}{2}=\eta\frac{Z-q(\tau)}{\Delta(\tau)},\quad \phi(Z,\tau)=\varphi(\tau),
	\end{equation}
	in which $q(\tau)$, $\varphi(\tau)$, and $\Delta(\tau)$ respectively
	indicate the center position, tilting angle, and width of the AFM-180DW.
	$\eta=+1$ or $-1$ represents a head-to-head (HH) or tail-to-tail (TT) wall.
	By letting $\Lambda$ take $q(\tau)$, $\varphi(\tau)$, and $\Delta(\tau)$ successively and integrating 
	over the long axis (i.e., $\int_{-\infty}^{+\infty}dZ$), Eq. (\ref{EulerLagrangeRayleighEquation}) provides
       \begin{subequations}\label{GeneralDynamicsEquations}
			\begin{eqnarray}
				0&=&\frac{\left(\ddot{q}+\alpha_{\mathrm{G}}\dot{q}\right)\Delta-\dot{q}\dot{\Delta}}{\Delta^2}-\eta\frac{\pi}{2}\frac{t_0}{t_{\parallel}}\sin\theta_{\mathrm{p}}\sin\left(\varphi-\phi_{\mathrm{p}}\right), \label{GDEa}
				\\
				0&=&\ddot{\varphi}+\frac{\dot{\varphi}\dot{\Delta}}{\Delta}+\frac{\kappa_y}{2}\sin2\varphi+\alpha_{\mathrm{G}}\dot{\varphi}-\frac{t_0}{t_{\parallel}}\cos\theta_{\mathrm{p}}, \label{GDEb}
				\\
				0&=& \frac{6}{\pi^2}\left[\frac{\dot{q}^2-1}{\Delta^2}-\dot{\varphi}^2+\kappa_y\sin^2\varphi+\kappa_z\right] 
				\nonumber \\
				&&\quad +\frac{\ddot{\Delta}+\alpha_{\mathrm{G}}\dot{\Delta}}{\Delta}-\frac{\dot{\Delta}^2}{2\Delta^2},  \label{GDEc}
			\end{eqnarray}
		\end{subequations}
	where a dot (two dots) over a variable means its first-order (second-order) partial derivative with 
	respect to $\tau$,
	$\theta_{\mathrm{p}}$ ($\phi_{\mathrm{p}}$) is the polar (azimuthal) angle of $\mathbf{p}_{\mathrm{cur}}$,
	and $(t_{\parallel})^{-1}\equiv\epsilon j_{\mathrm{cur}}a_0^3/(8Sde)$ denotes the spin-transfer rate.
	This set of equations is the central result of our analysis in this work. 
	
	Based on it, we explore three typical polarizers. 
	We start from the simplest case: parallel polarizers, i.e. $\theta_{\mathrm{p}}=0$ and arbitrary $\phi_{\mathrm{p}}$.
	Then Eqs. (\ref{GDEa}) and (\ref{GDEb}) become
	\begin{subequations}\label{ParallelPolarizer_DynamicsEquations}
		\begin{eqnarray}
			0&=&\left(\ddot{q}+\alpha_{\mathrm{G}}\dot{q}\right)\Delta-\dot{q}\dot{\Delta}, \label{DE_ParallelPolarizer_a}
			\\
			0&=&\ddot{\varphi}+\frac{\dot{\varphi}\dot{\Delta}}{\Delta}+\frac{\kappa_y}{2}\sin2\varphi+\alpha_{\mathrm{G}}\dot{\varphi}-\frac{t_0}{t_{\parallel}}, \label{DE_ParallelPolarizer_b}
		\end{eqnarray}
	\end{subequations}
	with unchanged Eq. (\ref{GDEc}). 
	We focus on the rigid-body mode, that is, $\dot{\varphi}_0=\ddot{\varphi}_0=0$, 
	$\dot{\Delta}_0=\ddot{\Delta}_0=0$, and $\ddot{q}_0=0$.
	This leads to
	\begin{equation}\label{ParallelPolarizer_results}
		\dot{q}_0=0, \quad \sin2\varphi_0=\frac{2t_0}{\kappa_y t_{\parallel}}, \quad \Delta_0=\frac{1}{\sqrt{\kappa_z+\kappa_y\sin^2\varphi_0}},
	\end{equation}
    implying an upper limit of current density,
    $j_{\mathrm{cur}}^{\mathrm{up}}\equiv4\kappa_y JS^2 de/(\epsilon\hbar a_0^3)$,
    for the existence of rigid-body mode.
    In addition, there are two possibilities for the tilting angle: 
    $\varphi_0^{\pm}=\arcsin\sqrt{[1\pm\sqrt{1-(2t_0)^2/(\kappa_y t_{\parallel})^2}]/2}$.
    To determine which one is stable, we set 
    \begin{subequations}\label{ThreeVariations}
    	\begin{eqnarray}
    		q&=&q_0+\delta q,\quad\  |\delta q|\ll 1 \label{ThreeVariations_a}
    		\\
    		\varphi&=&\varphi_0+\delta\varphi,\quad |\delta\varphi|\ll |\varphi_0|, \label{ThreeVariations_b}
    		\\
    		\Delta&=&\Delta_0+\delta\Delta,\quad |\delta\Delta|\ll |\Delta_0|.  \label{ThreeVariations_c}
    	\end{eqnarray}
    \end{subequations}
    By putting Eqs. (\ref{ParallelPolarizer_results}) and (\ref{ThreeVariations}) into Eq. (\ref{GeneralDynamicsEquations})
    and preserving only the linear terms of $\delta q$, $\delta\varphi$ and $\delta\Delta$, we have
    \begin{subequations}\label{ParallelPolarizer_ThreeVariationEquations}
    	\begin{eqnarray}
    		0&=& \frac{\partial^2(\delta q)}{\partial \tau^2} +\alpha_{\mathrm{G}}\frac{\partial (\delta q)}{\partial\tau}      \label{ParallelPolarizer_ThreeVariationEquations_a}
    		\\
    		0&=& \frac{\partial^2(\delta \varphi)}{\partial \tau^2} + \alpha_{\mathrm{G}}\frac{\partial (\delta \varphi)}{\partial\tau}  +\kappa_y\cos2\varphi_0\delta\varphi,   \label{ParallelPolarizer_ThreeVariationEquations_b}
    		\\
    		0&=& \frac{\partial^2(\delta \Delta)}{\partial \tau^2} +\alpha_{\mathrm{G}}\frac{\partial (\delta \Delta)}{\partial\tau} +\frac{6\Delta_0}{\pi^2}\left(\frac{2\delta\Delta}{\Delta_0^3}+\frac{2t_0}{t_{\parallel}}\delta\varphi\right).  \label{ParallelPolarizer_ThreeVariationEquations_c}
    	\end{eqnarray}
    \end{subequations}
    Therefore, only the $\varphi_0^{+}-$branch is stable.
    In summary, under parallel polarizers AFM-180DWs may creep for a while but will 
    eventually come to a standstill, as long as we do not consider the instability of walls under 
    extremely high $j_{\mathrm{cur}}$.
	
	Next we turn to the perpendicular polarizers ($\theta_{\mathrm{p}}=\pi/2$ and $\phi_{\mathrm{p}}=\pi/2$).
	Equations (\ref{GDEa}) and (\ref{GDEb}) now become
	\begin{subequations}\label{PerpendicularPolarizer_DynamicsEquations}
		\begin{eqnarray}
			0&=&\frac{\left(\ddot{q}+\alpha_{\mathrm{G}}\dot{q}\right)\Delta-\dot{q}\dot{\Delta}}{\Delta^2}+\eta\frac{\pi}{2}\frac{t_0}{t_{\parallel}}\cos\varphi, \label{DE_PerpendicularPolarizer_a}
			\\
			0&=&\ddot{\varphi}+\frac{\dot{\varphi}\dot{\Delta}}{\Delta}+\frac{\kappa_y}{2}\sin2\varphi+\alpha_{\mathrm{G}}\dot{\varphi}, \label{DE_PerpendicularPolarizer_b}
		\end{eqnarray}
	\end{subequations}
	meantime leaving Eq. (\ref{GDEc}) unchanged. 
	Again we focus on the rigid-body mode, which now has the following two branches: 
	the ``$k\pi$'' ($k\in \mathbb{Z}$) one with nonzero wall velocity 
	\begin{equation}\label{PerpendicularPolarizer_n_pi_branch}
		\dot{q}_1=\frac{\eta(-1)^{k+1}\xi}{\sqrt{\kappa_z+\xi^2}}, \quad \varphi_1=k\pi, \quad \Delta_1=\frac{1}{\sqrt{\kappa_z+\xi^2}},
	\end{equation}
    where $\xi\equiv \frac{\pi}{2\alpha_{\mathrm{G}}}\frac{t_0}{t_{\parallel}}$,
	and the ``$(k+1/2)\pi$'' one with zero velocity
	\begin{equation}\label{PerpendicularPolarizer_n_plus_half_pi_branch}
		\dot{q}_2=0, \quad \varphi_2=\left(k+\frac{1}{2}\right)\pi, \quad \Delta_2=\frac{1}{\sqrt{\kappa_z+\kappa_y}}.
	\end{equation}
	Stability analysis indicates that the ``$(k+1/2)\pi$'' branch is not guaranteed to be stable.
	While the ``$k\pi$'' branch will be stable when $(12/\pi^2)\kappa_z-\alpha_{\mathrm{G}}^2>0$, 
	which holds for nearly all realistic systems.
	This means that spin currents filtered by perpendicular polarizers can induce
	steady propagation of AFM-180DWs.
	In addition, several interesting inferences can be drawn from Eq. (\ref{PerpendicularPolarizer_n_pi_branch}):
	(i) The direction of wall motion depends on both the wall's topological charge ($\eta$) and	the tiling angle ($k$);
	(ii) The magnitude of wall velocity is positively correlated with $j_{\mathrm{cur}}$
	and approaches 1 (in dimensional form $v_{\mathrm{SW}}\equiv a_0/t_0$, which is the group velocity of AFM spin wave in long-wavelength limit\cite{Kittel_2005}) under high enough $j_{\mathrm{cur}}$.
	Of course in reality $j_{\mathrm{cur}}$ can not be too large, otherwise the wall would deform or even collapse.
	(iii) The wall velocity is negatively correlated with $\alpha_{\mathrm{G}}$, 
	similar	to that in FM spin valves.
	
	Then we arrive at the planar-transverse polarizers ($\theta_{\mathrm{p}}=\pi/2$, $\phi_{\mathrm{p}}=0$). 
	Similar algebra also provides two branches:
	the ``$k\pi$'' branch with zero wall velocity
	\begin{equation}\label{PlanarTransversePolarizer_n_pi_branch}
		\dot{q}_3=0, \quad \varphi_3=k\pi, \quad \Delta_3=\frac{1}{\sqrt{\kappa_z}},
	\end{equation}
    and the ``$(k+1/2)\pi$'' branch with nonzero velocity
	\begin{equation}\label{PlanarTransversePolarizer_n_plus_half_pi_branch}
	\dot{q}_4=\eta(-1)^{k}\xi\Delta_4, \ \  \varphi_4=\left(k+\frac{1}{2}\right)\pi, \ \  \Delta_4=\frac{1}{\sqrt{\kappa_z+\kappa_y+\xi^2}}.
	\end{equation}
	Only the ``$k\pi$'' branch	is stable, hence reduces the availability of reversing N\'{e}el vectors by planar-transverse polarizers.

	\begin{figure}
	\includegraphics[width=0.40\textwidth]{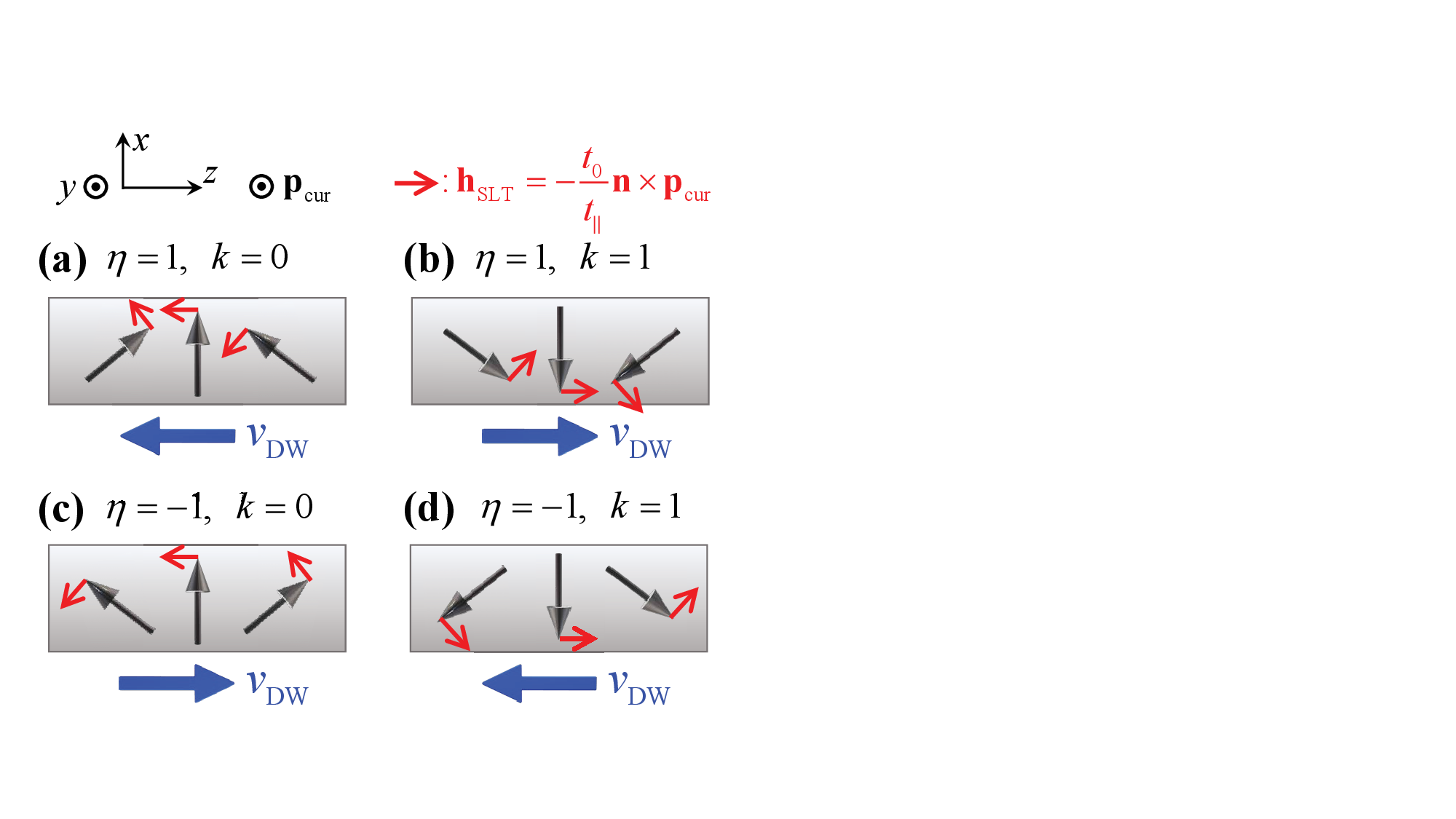}
	\caption{\label{fig2} Illustration of the dynamics of four typical AFM-180DWs ($\eta=\pm1$, $k=\pm1$) under perpendicular polarizers. The black, red and blue arrows respectively indicate
	the N\'{e}el vectors, the Slonczewski effective fields, and the propagation directions of walls.}
    \end{figure}
	
	In a word, our analytics has proved that for HH or TT AFM-180DWs, 
	only spin currents stimulated by perpendicular polarizers can drive them to propagate steadily along the AFM strip.
	This can be understood as follows.
	By setting $\mu=Z,\tau$, $\Lambda=\mathbf{n}$ and taking into account
	$|\mathbf{n}|\equiv1$, Eq. (\ref{EulerLagrangeRayleighEquation}) provides
	\begin{eqnarray}\label{EulerLagrangeRayleighEquation_vectorial}
			\frac{\partial^2\mathbf{n}}{\partial\tau^2}\times\mathbf{n}&=&\frac{\partial^2\mathbf{n}}{\partial Z^2}\times\mathbf{n}-\kappa_y n_y \mathbf{e}_y\times\mathbf{n}+\kappa_z n_z \mathbf{e}_z\times\mathbf{n}         \nonumber	\\
		     & & +\alpha_{\mathrm{G}}\mathbf{n}\times\frac{\partial\mathbf{n}}{\partial\tau}+\frac{t_0}{t_{\parallel}}\mathbf{n}\times\left(\mathbf{n}\times\mathbf{p}_{\mathrm{cur}}\right).
	\end{eqnarray}
	The first three terms on the right hand can be regrouped to $\mathbf{n}\times\mathbf{h}_{\mathrm{eff}}$
	with $\mathbf{h}_{\mathrm{eff}}\equiv-\frac{\partial^2\mathbf{n}}{\partial Z^2}+\kappa_y n_y \mathbf{e}_y-\kappa_z n_z \mathbf{e}_z$ describing the internal exchange and anisotropy contributions.
	The fourth term indicates the damping process. 
	The last term is often called the ``Slonczewski torque'' with the corresponding effective field
	$\mathbf{h}_{\mathrm{SLT}}\equiv -\frac{t_0}{t_{\parallel}}\mathbf{n}\times\mathbf{p}_{\mathrm{cur}}$ 
	describing the effect of vertically injected spin current on the N\'{e}el vector.
	In Fig. \ref{fig2} we illustrate four possibilities ($\eta=\pm1$, $k=\pm1$) of a AFM-180DW 
	lying in the easy plane under perpendicular polarizers.
	The black arrows represent the N\'{e}el vectors around the wall centers.
	After injecting the vertical spin current, the Slonczewski effective fields $\mathbf{h}_{\mathrm{SLT}}$ emerge and
	are indicated by the red arrows at each position of N\'{e}el vectors.
	The damping process will cause the N\'{e}el vector to rotate towards $\mathbf{h}_{\mathrm{SLT}}$, leading to the 
	propagation of walls along $Z$ axis.
	Note that for $\eta=1$ ($-1$) the wall motion and rotation of N\'{e}el vector at wall center are homodromous (heterodromous).
	Alternatively, if the wall lies in $YZ$ plane, its energy is at the saddle point rather than a minimum.
	The corresponding Slonczewski effective field only results in a twisting (thus instability) of the wall but no driving effect.
	Similar discussions apply to ``planar-transverse polarizer'' case and we will not elaborate further.

	Before the end of this letter, we perform typical numerics to test our analytics.
	The magnetic parameters are as follows.
	The geometry of our AFM strip is $w=100$ nm, $d=5$ nm and $L_z=10\ \mu \mathrm{m}$ 
	with $\alpha_{\mathrm{G}}=0.01$.
	The bond length $c=1$ nm in all directions, thus the unit cell has the dimension 
	$(a_0/2) \times (a_0/2) \times a_0$ with $a_0=2$ nm, and $2N=10^4$.
	Its exchange field is $\mu_0 H_E=100$ T, leading to $t_0=(\gamma_0 H_E)^{-1}=5.68\times 10^{-14}$ s
	and $J=\hbar/(t_0 S)=1.856\times10^{-21}$ J (by setting $S=1$).
	The biaxial anisotropy fields read $\mu_0 H_K^z=1$ T (easy) and $\mu_0 H_K^y=0.05$ T (hard),
	hence $\kappa_z=8K_z/J=0.01$ and $\kappa_y=8K_y/J=5\times 10^{-4}$.
	By setting $\epsilon=0.5$ we have $t_0/t_{\parallel}=3.547\times 10^{-10} j_{\mathrm{cur}}$, 
	in which $j_{\mathrm{cur}}$ takes the unit of ``$\mathrm{A/cm^2}$''.
	Finally, the upper limit of wall velocity is $v_{\mathrm{SW}}=3.52\times 10^4 \ \mathrm{m/s}$,
	leading to a nanosecond-level (Terahertz) reversal of N\'{e}el vectors.
	
	Our numerics are performed at two levels.
	At the first one, we check whether our rigid-body results coincide with the long-term stable solutions of 
	Eq. (\ref{GeneralDynamicsEquations}), which is obtained via
	the fourth-order Runge-Kutta algorithm.
	At the second level, we perform numerical simulations based on the discrete lattice model 
	in Eq. (\ref{HeisenbergHamiltonianDiscrete}).
	The two unit vectors along the respective magnetic moments in the ``$i-$''th unit cell, that is $\mathbf{m}_{i}^{A(B)}$, 
	satisfy the following LLG-Slonczewski equation set
	\begin{widetext}
	\begin{eqnarray}\label{LLG_Slonczewski_Equation_Set}
		\frac{\partial\mathbf{m}_{i}^{A(B)}}{\partial\tau}&=&-\mathbf{m}_{i}^{A(B)}\times\mathbf{h}_{i}^{A(B)} +\frac{\alpha_{\mathrm{G}}}{4}\mathbf{m}_{i}^{A(B)}\times\frac{\partial\mathbf{m}_{i}^{A(B)}}{\partial\tau}  
        +\frac{t_0}{4t_{\parallel}}\mathbf{m}_{i}^{A(B)}\times\left(\mathbf{m}_{i}^{A(B)}\times\mathbf{p}_{\mathrm{cur}}\right),  \nonumber  \\
		\mathbf{h}_{i}^{A}&=& 
		-\left(\mathbf{m}_{i-1}^{B}+\mathbf{m}_{i}^{B}\right)-\frac{\kappa_y}{4}\left(\mathbf{m}_{i}^{A}\cdot\mathbf{e}_y\right)\mathbf{e}_y + \frac{\kappa_z}{4}\left(\mathbf{m}_{i}^{A}\cdot\mathbf{e}_z\right)\mathbf{e}_z,                \nonumber   \\
		\mathbf{h}_{i}^{B}&=&
		-\left(\mathbf{m}_{i}^{A}+\mathbf{m}_{i+1}^{A}\right)-\frac{\kappa_y}{4}\left(\mathbf{m}_{i}^{B}\cdot\mathbf{e}_y\right)\mathbf{e}_y + \frac{\kappa_z}{4}\left(\mathbf{m}_{i}^{B}\cdot\mathbf{e}_z\right)\mathbf{e}_z,
	\end{eqnarray}
    \end{widetext}
	where $1\le i \le N$ and $\mathbf{m}_{0}^{B}=\mathbf{m}_{N+1}^{A}\equiv 0$.
	Equations (\ref{LLG_Slonczewski_Equation_Set}) and (\ref{EulerLagrangeRayleighEquation_vectorial}) 
	are equivalent after ignoring the second and higher orders of $|\mathbf{m}_i|$ (or $|\mathbf{m}|$).

	\begin{figure}
	\includegraphics[width=0.45\textwidth]{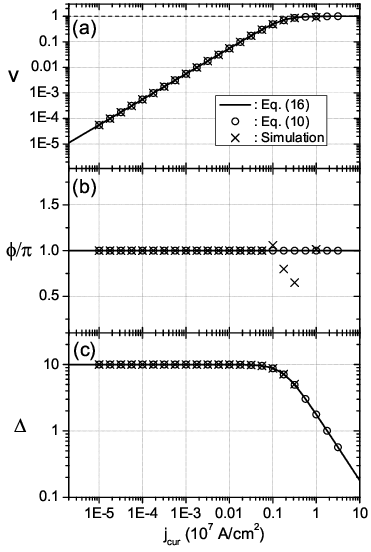}
	\caption{\label{fig3} Comparison between analytics and numerics under
	perpendicular polarizers for walls in Fig. \ref{fig2}(b). The solid curves come from 
	Eq. (\ref{PerpendicularPolarizer_n_pi_branch}). The open circles are the long-term stable solutions of
	Eq. (\ref{GeneralDynamicsEquations}). The cross stars denote simulation results 
	based on Eq. (\ref{LLG_Slonczewski_Equation_Set}).}
    \end{figure}
	
	In Fig. \ref{fig3}, we plot the $j_{\mathrm{cur}}-$dependence of the velocity, tilting angle and width of a AFM-180DW
	in Fig. \ref{fig2}(b) under perpendicular polarizers.
	The solid curves, hollow circles and cross stars are data respectively from the analytical results in 
	Eq. (\ref{PerpendicularPolarizer_n_pi_branch}), long-term stable solutions of 
	Eq. (\ref{GeneralDynamicsEquations}), and numerical simulations 
	based on Eq. (\ref{LLG_Slonczewski_Equation_Set}).
    Note that tilting angles from simulations are the average ones across the whole wall.
	All three groups of data coincide perfectly, indicating the correctness of our analytics.
	It should be noted that in simulations when $J_{\mathrm{cur}} \ge 1.0\times 10^{6}\ \mathrm{A/cm^2}$,
	the stronger Slonczewski torque make the $\varphi-$plane suffer a distortion but the wall velocity remains regular
	and approaches $v_{\mathrm{SW}}$ when $J_{\mathrm{cur}} \rightarrow 1.0\times 10^{7}\ \mathrm{A/cm^2}$.
	Furthermore, for $J_{\mathrm{cur}} > 1.0\times 10^{7}\ \mathrm{A/cm^2}$ AFM-180DWs begin to distort severely or 
	even collapse, hence we can not acquire reasonable data for wall width and velocity from simulations.
	More interestingly, under $J_{\mathrm{cur}}=5.5\times 10^{4}\ \mathrm{A/cm^2}$ ($5.5\times 10^{5}\ \mathrm{A/cm^2}$), AFM-180DWs can acquire approximately 3\% (30\%) of $v_{\mathrm{SW}}$ that already exceed $10^3$ ($10^4$) m/s. 
	This greatly improves the practicability of our proposal under small current densities.

	In summary, we have revealed a new possibility of achieving ultrafast propagation of AFM-180DWs
	in bipartite AFM strips (either conducting or insulating) by vertically injected spin currents.
	Among three typical polarizations the spin current would have, only the perpendicular ones 
	can sustain a steady rigid-body propagation with high-enough velocity.
	This has been first theoretically proposed, then numerically confirmed under a practical set 
	of magnetic parameters.
	Our results should inspire the development of novel magnetic nanodevices based on fine manipulations of 
	N\'{e}el vectors in AFMs, especially the insulating ones with excellent properties.

		\begin{acknowledgments}
			M.L. acknowledges support from the National Natural Science Foundation of China (Grant No. 12204403).
			J.L. is supported by the Natural Science Foundation of Jiangsu Province (Grant No. BK20241929).
		\end{acknowledgments}
	
	    \section*{Author Declarations}
	    \subsection*{Conflict of Interest}
	    The authors have no conflicts to disclose.
	    
	    \subsection*{Author Contributions}
		\textbf{Jiaxin Du}: Investigation (equal); Methodology (equal); 
		                             Visualization (equal); Writing -- original draft (equal). 
		\textbf{Mei Li}: Funding acquisition (equal); Data curation (equal); Investigation (equal); Methodology (equal); 
		                         Visualization (equal). 
		\textbf{Bin Xi}: Data curation (equal); Investigation (equal); Methodology (equal); Visualization (equal). 
		\textbf{Xue Zhang}: Investigation (equal); Visualization (equal). 
		\textbf{Yi Sun}: Investigation (equal); Visualization (equal). 
		\textbf{Chun-Gui Duan}: Conceptualization (equal); Investigation (equal); Supervision (equal); Writing -- original draft (equal). 
		\textbf{Jie Lu}: Conceptualization (equal); Formal analysis; Funding acquisition (equal); Investigation (equal); 
		Project administration; Supervision (equal); Writing -- original draft (equal); Writing -- review \& editing.
		
		\section*{Data Availability}
		
		The data that support the findings of this study are available from the corresponding author upon reasonable request.


\section*{References}
\begin{thebibliography}{999}
\bibitem{Tserkovnyak_RMP_2018} V. Baltz, A. Manchon, M. Tsoi, T. Moriyama, T. Ono, and Y. Tserkovnyak, ``Antiferromagnetic spintronics'', Rev. Mod. Phys. \textbf{90}, 015005 (2018).
\bibitem{XRWang_EPL_2020} H. Y. Yuan, Z. Yuan, R. A. Duine, and X. R. Wang, ``Recent progress in antiferromagnetic dynamics'', Europhy. Lett., \textbf{132}, 57001 (2020).
\bibitem{HaimingYu_PhysRep_2021} H. Yu, J. Xiao, and H. Schultheiss, ``Magnetic texture based magnonics'', Phys. Rep. \textbf{905}, 1 (2021).
\bibitem{LiuZhiqi_AM_2024} H. Chen, L. Liu, X. Zhou, Z. Meng, X. Wang, Z. Duan, G. Zhao, H. Yan, P. Qin, and Z. Liu, ``Emerging Antiferromagnets for Spintronics'', Adv. Mater.  \textbf{36}, 2310379 (2024).

\bibitem{RCheng_PRL_2016} R. Cheng, D. Xiao, and A. Brataas, ``Terahertz antiferromagnetic spin hall nano-oscillator'', Phys. Rev. Lett. \textbf{116}, 207603 (2016).
\bibitem{Chirac_PRB_2020} T. Chirac, J.-Y. Chauleau, P. Thibaudeau, O. Gomonay, and M. Viret, ``Ultrafast antiferromagnetic switching in NiO induced by spin transfer torques'', Phys. Rev. B \textbf{102}, 134415 (2020).


\bibitem{Shadman_APL_2019} A. Shadman and J.-G. (Jimmy) Zhu, ``High-speed STT MRAM incorporating antiferromagnetic layer'', Appl. Phys. Lett. \textbf{114}, 022403 (2019).

\bibitem{Fischer_PRApp_2020} J. Fischer, M. Althammer, N. Vlietstra, H. Huebl, S. T. B. Goennenwein, R. Gross, S. Gepr\"{a}gs, and M. Opel, ``Large spin Hall magnetoresistance in antiferromagnetic $\alpha-\mathrm{Fe_2 O_3}$/Pt heterostructures'', Phys. Rev. Appl. \textbf{13}, 014019 (2020).



\bibitem{HaoWu_NC_2022} H. Wu, H. Zhang, B. Wang, F. Gro\ss, C.-Y. Yang, G. Li, C. Guo, H. He, K. W, D. Wu, X. Han, C.-H. Lai, J. Gr\"{a}fe, R. Cheng, and K. L. Wang, ``Current-induced N\'{e}el order switching facilitated by magnetic phase transition'', Nat. Commun. \textbf{13}, 1629 (2022).
\bibitem{Weissenhofer_PRB_2023} M. Wei\ss enhofer, F. Foggetti, U. Nowak, and P. M. Oppeneer, ``N\'{e}el vector switching and terahertz spin-wave excitation in $\mathrm{Mn_2 Au}$ due to femtosecond spin-transfer torques'', Phys. Rev. B \textbf{107}, 174424 (2023).
\bibitem{SongCheng_SciAdv_2024}  Y. Zhou, T. Guo, L. Han, L. Liao, W. He, C. Wan, C. Chen, Q. Wang, L. Qiao, H. Bai, W. Zhu, Y. Zhang, R. Chen, X. Han, F. Pan, and C. Song, ``Spin-torque–driven antiferromagnetic resonance'', Sci. Adv. \textbf{10}, eadk7935 (2024).



\bibitem{XRWang_PRB_2018} H. Y. Yuan, W. Wang, M.-H. Yung, and X. R. Wang, ``Classification of magnetic forces acting on an antiferromagnetic domain wall'', Phys. Rev. B \textbf{97}, 214434 (2018).

\bibitem{Tchernyshyov_PRB_2014} S. K. Kim, Y. Tserkovnyak, and O. Tchernyshyov, ``Propulsion of a domain wall in an antiferromagnet by magnons'', Phys. Rev. B \textbf{90}, 104406 (2014).
\bibitem{Brataas_PRB_2018} A. Qaiumzadeh, L. A. Kristiansen, and A. Brataas, ``Controlling chiral domain walls in antiferromagnets using spin-wave helicity'', Phys. Rev. B \textbf{97}, 020402(R) (2018).


\bibitem{NiuQian_PRB_2014} R. Cheng and Q. Niu, ``Dynamics of antiferromagnets driven by spin current'', Phys. Rev. B \textbf{89}, 081105(R) (2014).
\bibitem{Sinova_PRB_2016} Y. Yamane, J. Ieda, and J. Sinova, ``Spin-transfer torques in antiferromagnetic textures: Efficiency and quantification method'', Phys. Rev. B \textbf{94}, 054409 (2016).
\bibitem{Rodrigues_PRB_2017} D. R. Rodrigues, K. Everschor-Sitte, O. A. Tretiakov, J. Sinova, and Ar. Abanov, ``Spin texture motion in antiferromagnetic and ferromagnetic nanowires'', Phys. Rev. B \textbf{95}, 174408 (2017).
\bibitem{Mondal_PRB_2018} R. Mondal, M. Berritta, and P. M. Oppeneer, ``Unified theory of magnetization dynamics with relativistic and nonrelativistic spin torques'', Phys. Rev. B \textbf{98}, 214429 (2018).
\bibitem{Fujimoto_PRB_2021} J. Fujimoto, ``Adiabatic and nonadiabatic spin-transfer torques in antiferromagnets'', Phys. Rev. B \textbf{103}, 014436 (2021).
\bibitem{XRWang_PRB_2024} K. Y. Jing, Z.-Z. Sun, and X. R. Wang, ``Current-driven domain wall motion in ferrimagnetic nanowires'', Phys. Rev. B \textbf{110}, 054414 (2024).

\bibitem{Rebei_Mryasov_PRB_2006} A. Rebei and O. Mryasov, ``Dynamics of a trapped domain wall in a spin-valve nanostructure with current perpendicular to the plane'', Phys. Rev. B \textbf{74}, 014412 (2006).
\bibitem{Khvalkovskiy_PRL_2009} A. V. Khvalkovskiy,, K. A. Zvezdin, Ya. V. Gorbunov, V. Cros, J. Grollier, A. Fert, and A. K. Zvezdin, ``High domain wall velocities due to spin currents perpendicular to the plane'', Phys. Rev. Lett. \textbf{102}, 067206 (2009).
\bibitem{Boone_PRL_2010} C. T. Boone, J. A. Katine, M. Carey, J. R. Childress, X. Cheng, and I. N. Krivorotov, ``'', Phys. Rev. Lett. \textbf{104}, 097203 (2010).
\bibitem{PBHe_EPJB_2013} P.-B. He, ``Current-driven domain wall dynamics in spin-valve nanostrips with parallel, perpendicular, and tilted polarizers'', Eur. Phys. J. B \textbf{86}, 412 (2013).
\bibitem{jlu_PRB_2019} M. Li, Z. An, and J. Lu, ``Ultrahigh differential mobility and velocity of N\'{e}el domain walls in spin valves with planar-transverse polarizers and perpendicularly injected small currents'', Phys. Rev. B \textbf{100}, 064406 (2019).
\bibitem{jlu_PRB_2021} J. Du, M. Li, and J. Lu, ``Stabilizing current-driven steady flows of 180$^{\circ}$ domain walls in spin valves by interfacial Dzyaloshinskii-Moriya interaction'', Phys. Rev. B \textbf{103}, 144429 (2021).
\bibitem{jlu_NJP_2023} J. Du, M. Li, X. Zhang, B. Xi, Y.-J. Liu, C.-G. Duan, and J. Lu, ``Slonczewski-spin-current driven dynamics of 180$^{\circ}$ domain walls in spin valves with interfacial Dzyaloshinskii–Moriya interaction'', New J. Phys. \textbf{25}, 093050 (2023).

\bibitem{Ralph_JMMM_2008} D.C. Ralph and M.D. Stiles, ``Spin transfer torques'', J. Magn. Magn. Mater. \textbf{320}, 1190 (2008).
\bibitem{Gomonay_LowTempPhys_2014} E. V. Gomonay and V. M. Loktev, ``Spintronics of antiferromagnetic systems (Review Article)'', Low Temp. Phys. \textbf{40}, 17 (2014).
\bibitem{KWKim_IEEENano_2020} Y. G. Semenov and K. W. Kim, ``Modeling of antiferromagnetic dynamics: a brief review'', IEEE Nanotechnology Magazine \textbf{14}, 32 (2020).

\bibitem{MacDonald_PRL_2008} P. M. Haney and A. H. MacDonald, ``Current-induced torques due to compensated antiferromagnets'',  Phys. Rev. Lett. \textbf{100}, 196801 (2008).
\bibitem{ZhenWei_JAP_2009} Z. Wei, J. Basset, A. Sharma, J. Bass, and M. Tsoi, ``Spin-transfer interactions in exchange-biased spin valves'', J. Appl. Phys. \textbf{105}, 07D108 (2009).
\bibitem{Haney_PRB_2014} K. Prakhya, A. Popescu, and P. M. Haney, ``Current-induced torques between ferromagnets and compensated antiferromagnets: Symmetry and phase coherence effects'', Phys. Rev. B \textbf{89}, 054421 (2014).
\bibitem{Gomonay_LowTempPhys_2008} E. V. Gomonay and V. M. Loktev, ``Distinctive effects of a spin-polarized current on the static and dynamic properties of an antiferromagnetic conductor'', Low Temp. Phys. \textbf{34}, 198 (2008).
\bibitem{Gomonay_PRB_2010} E. V. Gomonay and V. M. Loktev, ``Spin transfer and current-induced switching in antiferromagnets'', Phys. Rev. B \textbf{81}, 144427 (2010).
\bibitem{Gomonay_PRB_2012} H. V. Gomonay, R. V. Kunitsyn, and V. M. Loktev, ``Symmetry and the macroscopic dynamics of antiferromagnetic materials in the presence of spin-polarized current'', Phys. Rev. B \textbf{85}, 134446 (2012).
\bibitem{KWKim_APL_2017} Y. G. Semenov and K. W. Kim, ``Spin pumping torque in antiferromagnets'', Appl. Phys. Lett. \textbf{110}, 192405 (2017).
\bibitem{KWKim_PRB_2021} Y. G. Semenov and K. W. Kim, ``Spin-transfer and fieldlike torques in antiferromagnets'', Phys. Rev. B \textbf{104}, 174402 (2021).


\bibitem{Brataas_PRB_2016} E. G. Tveten, T. M\"{u}ller, J. Linder, and A. Brataas, ``Intrinsic magnetization of antiferromagnetic textures'', Phys. Rev. B \textbf{93}, 104408 (2016).
\bibitem{Stamps_PRB_2021} J. O. Iyaro, I. Proskurin, and R. L. Stamps, ``Collective dynamics of domain walls: An antiferromagnetic spin texture in an optical cavity'', Phys. Rev. B \textbf{104}, 184416 (2021).


\bibitem{Kittel_2005} C. Kittel, Introduction to Solid State Physics, 8th Edition (John Wiley \& Sons, 2005), pp 345.








	
\end{thebibliography}
	\end{document}